\documentstyle[11pt,IAUS212,twoside,epsf]{article}

\def\edcomment#1{\iffalse\marginpar{\raggedright\sl#1\/}\else\relax\fi}
\reversemarginpar

\begin{document}
\vspace*{1cm}
\title{Searching for Hydrogen in Type~Ib Supernovae, and for Helium and
Hydrogen in Type~Ic }
\author{David Branch}
\affil{Department of Physics and Astronomy, University of
Oklahoma, Norman, Oklahoma 73019, USA}

\begin{abstract}
Identifying the progenitors of Type~Ib and Type~Ic supernovae requires
knowing, among other things, whether SNe~Ib eject hydrogen, and
whether SNe~Ic eject helium, and perhaps even hydrogen.  Recently it
has become clear that some SNe~Ib do eject hydrogen, and it may be
that all SNe~Ib do.  Two arguments that have been made in the past
that SNe~Ic eject helium are difficult to confirm, but I discuss other
possible evidence that SNe~Ic eject helium, as well as hydrogen.  If
so, these elements extend to {\sl lower} ejection velocities than in
SNe~Ib.  The spectroscopic differences between SNe~Ib and SNe~Ic may
depend on the radial distributions of the helium and hydrogen as well
as on the ejected masses of helium and hydrogen.  We should consider
the possibility that SNe~Ic are more mixed up.
\end{abstract}

\section{Introduction}

Type~Ia supernovae are thought to be nuclear--powered disruptions of
accreting or merging white dwarfs.  Most if not all other supernovae
(SNe) are thought to involve expulsion of the envelopes of massive
stars following the gravitational collapse of their highly evolved
cores.  Of these, Type~II SNe have conspicuous features due to
hydrogen in their optical spectra.  Type~Ib SNe lack conspicuous
hydrogen features but have conspicuous He~I features during their
photospheric phase.  In the spectra of Type~Ic SNe, neither hydrogen
nor He~I features are conspicuous.  For reviews of SN classification
see Filippenko (1997) and Turatto (2002).

The optical spectrum of a typical SN~II evolves from a nearly
featureless continuum when the temperature is high to one that
contains mainly hydrogen features and then gradually develops features
of lower excitation due to Ca~II, Fe~II, O~I, Na~I, and Ti~II, plus
Ba~II and Sr~II if the temperature gets sufficiently low.  Apart from
the defining differences involving the hydrogen and He~I features, the
spectra of SNe~Ib and SNe~Ic evolve in much the same way.

Among the questions that those who make explosion models of SNe~Ib and
SNe~Ic would like spectroscopists to answer include two very basic
ones.  Do SNe~Ib eject any hydrogen?  And do SNe~Ic eject any helium
--- and perhaps even some hydrogen?  I will be able to provide only
partial and tentative answers to these questions --- but I should at
least be able to explain why they are so hard to answer.

\vfill\eject

\section{Direct Analysis of Supernova Spectra}

First we must say something about the analysis of SN spectra.  One
nice thing is that the simple explosion velocity law --- velocity
proportional to radius --- ordinarily is an excellent approximation.
One big problem is that ejection velocities usually are high, so
spectral features are blended and the amount of information that can
be obtained by analyzing the spectra with a ruler is limited.
Consequently, much use is made of synthetic spectra.  The use of
synthetic spectra for SNe has been reviewed by Branch, Baron, \&
Jeffery (2002).  Detailed and physically self--consistent NLTE
synthetic spectra can and have been carried out with very complex
codes, notably the PHOENIX code, but so far nearly all such work has
been for SNe~Ia and SNe~II.  At the other extreme, of simplicity, the
fast parameterized SYNOW code is useful for establishing line
identifications and placing constraints on the velocity intervals in
which the line--contributing ions are detected.  We refer to this
process as ``direct analysis'' to distinguish it from detailed
spectrum synthesis with more complicated codes.  The synthetic spectra
shown in this article have been generated with SYNOW.

SYNOW assumes spherical symmetry, velocity proportional to radius, and
a sharp photosphere that emits a blackbody continuum.  Line formation,
treated in the Sobolev approximation, occurs only by resonance
scattering of photons from the photosphere.  The effects of multiple
scattering (line blending) on the line source functions and the
emergent spectrum are treated exactly, within the context of the model
and the Sobolev approximation.  In the synthetic spectra shown in this
article, line optical depths are taken to decrease as $v^{-8}$.  For
each ion whose lines are introduced, the optical depth of a
``reference line'' at the photosphere is a fitting parameter, and the
optical depths of the other lines of the ion are determined by
assuming LTE level populations at some excitation temperature
$T_{exc}$, which for the synthetic spectra of this article is 7000~K.
Oscillator strengths are Kurucz (1993).  Line formation ordinarily
takes place immediately above the photosphere, i.e., the lower
boundary of the line--forming layer has the velocity at the
photosphere, $v_{phot}$.  But sometimes it is appropriate to
``detach'' the lines of an ion by allowing line formation only above
some detachment velocity that exceeds $v_{phot}$.

An undetached resonance--scattering line has a P~Cygni profile
consisting of a rounded emission component that peaks at the rest
wavelength, and an absorption component whose minimum is blueshifted
by the amount that corresponds to $v_{phot}$ (or more if the line is
strong).  A detached resonance--scattering line also has a P~Cygni
profile but the emission component has a flat top, and the absorption
minimum is blueshifted by the amount that corresponds to the
detachment velocity.  The absorption component of a detached line is
deeper than that of an undetached line that has the same maximum
Sobolev optical depth (Jeffery \& Branch 1990; Branch et~al. 2002).

\section{SNe~IIb}

Before discussing SNe~Ib we must mention SNe~IIb --- events in which
hydrogen features are conspicuous at early times but later become weak
or undetectable.  The first well observed example was SN~1993J in M81
(Matheson et~al. 2000 and numerous references therein). Conspicuous
hydrogen features were present at early times, but conspicuous He~I
features began to appear around 25 days after explosion; later the
hydrogen features became weak.  The minimum velocity of the
hydrogen--containing layer was about 9000 km~s$^{-1}$.  Most estimates
of the mass of this layer have been in the range 0.1 to 0.6~M$_\odot$;
the hydrogen mass was lower because much of the mass of this layer
probably was in the form of helium.  In SN~1996cb, the second well
observed SN~IIb (Qiu et~al. 1999), He~I features appeared earlier,
before maximum light, but a deep H$\alpha$ absorption persisted
throughout the photospheric phase.  The minimum velocity and mass of
the hydrogen--containing layer were about 10,000 km~s$^{-1}$ and
perhaps $\sim0.1$~M$_\odot$ (Deng, Qiu,\& Hu, 2002).

\section{Hydrogen in SNe~Ib?}

Until recently not much work had been done on searching for hydrogen
in SNe~Ib.  Branch (1972) considered H$\alpha$, as well as the
strongest optical line of Ne~I, $\lambda$6402, as possible identifications
of a deep absorption in SN~1954A. Wheeler et~al. (1994) tentatively
attributed a weak absorption in SNe~1983N and 1984L to H$\alpha$, as
did Deng et~al. (2000) in SN~1999dn.  An opportunity to study this
issue more systematically was presented by Matheson et~al. (2001,
hereafter M01) who published spectra of a number of SNe~Ib (and
SNe~Ic) that were observed at the Lick Observatory during the 1990s.

Recently the SYNOW code has been used to study the spectra of 11
SNe~Ib selected on the basis of having deep optical He~I absorptions
(Branch et~al. 2002, hereafter B02).  As had been emphasized by M01,
from whom much of the data of the B02 sample were taken, in most
respects the spectra of these events were quite homogeneous.  The
velocity at the photosphere, as inferred from Fe~II features near
5000~\AA, decreased smoothly from about 11,000 km~s$^{-1}$ at the time
of maximum light to about 5000 km~s$^{-1}$ 50 days later, with little
scatter. In postmaximum spectra the He~I absorptions generally were
detached, with the detachment velocity ending up at about 7000
km~s$^{-1}$.

Two events in the B02 sample, SNe~1999di and 2000H, differed from the
others by containing deep, persistent absorptions near 6300~\AA\ that
resembled the deep H$\alpha$ absorption in the Type~IIb SN~1996cb.  It
is safe to identify the 6300~\AA\ absorptions in these two SNe~Ib as
H$\alpha$, detached at 12,000 or 13,000 km~s$^{-1}$, because there is
no plausible alternative identification that would not entail unwanted
features in the spectrum, and because weak absorptions consistent with
H$\beta$ could be seen.  SN~1954A, for which of course only
photographic spectra are available, appears to have been similar. I
will refer to these events as ``deep--H$\alpha$'' SNe~Ib.  They are
classified as Type~Ib because when they were first observed the
presence of hydrogen features was not obvious, even though their
H$\alpha$ absorptions were deep.  The explanation, illustrated in
Figure~24 of B02, is that because the hydrogen lines were detached,
the H$\alpha$ emission components that we are accustomed to seeing in
SNe~II were flat and therefore inconspicuous, and the deep H$\alpha$
absorptions were produced with sufficiently small H$\alpha$ optical
depths that the other Balmer lines were too weak to be conspicuous.
(The Sobolev optical depth of H$\beta$ is lower than that of H$\alpha$
by a factor of 7.)  If these events had been observed earlier their
hydrogen features might have been stronger, undetached, and
conspicuous, as in the earliest spectra of SN~1996cb, in which case
these events would have been classified as SNe~IIb.

Among the eight other events of the B02 sample, four (SNe~1983N,
1984L, 1996N, and 1999dn) were first observed earlier than 10 days
after maximum light and they also had absorptions near 6300~\AA, but
less deep than in the deep--H$\alpha$ events. I will refer to these as
typical SNe~Ib.  In Figure~1 a spectrum of SN~1984L is compared with a
synthetic spectrum.  B02 attributed the 6300~\AA\ absorption in
typical SNe~Ib to H$\alpha$, detached at 18,000 km~s$^{-1}$ in the
earliest available spectrum and at 12,000 or 13,000 km~s$^{-1}$ when
it was last observed.  The identification was tentative, however,
because H$\beta$ could not be seen, and there are other possible line
identifications for the 6300~\AA\ absorption.  The strongest optical
line of C~II, $\lambda$6580, would need to be detached only 800
km~s$^{-1}$ more than H$\alpha$, but highly detached C~II lines would
be surprising in SNe~Ib. The strongest optical line of Si~II,
$\lambda$6355, would need to be blueshifted 9500 km~s$^{-1}$ less than
H$\alpha$, which is a problem because it would require the Si~II
feature to form in deeper layers than the Fe~II features.  A more
plausible alternative to H$\alpha$ is Ne~I $\lambda$6402, which would
need to be blueshifted 7400 km~s$^{-1}$ less than H$\alpha$; this
would put it right at the same velocity as the Fe~II lines.  The Ne~I
lines would need to be nonthermally excited, as are the He~I lines of
SNe~Ib.  When Fe~II features become strong they also can produce a
feature that could be mistaken for highly blueshifted H$\alpha$ (see
below), but in these four SNe~Ib the Fe~II features were not strong
enough around the time of maximum.  The best choices for the 6300~\AA\
absorption in typical SNe~Ib seem to be detached H$\alpha$ or
undetached Ne~I $\lambda$6402.

Only very recently has it become clear that some events have optical
He~I absorptions that are definite but not as deep as in typical
SNe~Ib.  I will refer to these as ``shallow--helium'' SNe~Ib.  One of
these, SN~1991D (Benetti et~al. 2002), deviates from typical SNe~Ib in
other respects too.  It was exceptionally luminous for a SN~Ib; 10
days after maximum light the velocity at the photosphere inferred from
the Fe~II features was only 5000 km~s$^{-1}$ as opposed to 8000
km~s$^{-1}$ for typical SNe~Ib; and at the same epoch the He~I lines
were not detached from the 5000 km~s$^{-1}$ photosphere whereas in
typical SNe~Ib they are detached at about 10,000 km~s$^{-1}$.
SN~1991D contained a weak 6300~\AA\ absorption that can be accounted
for by H$\alpha$ detached at 12,000 km~s$^{-1}$, but in SYNOW
synthetic spectra not only did Ne~I $\lambda$6402 provide a nice fit
to this feature, other Ne~I lines did more good than harm (Figure~3 of
Benetti et~al.).  Ne~I $\lambda$6402 appears to be a serious
alternative to H$\alpha$ in SN~1991D.

Hamuy et~al. (2002) have presented spectra of another shallow--helium
SN~Ib, SN~1999ex.  In this case the Fe~II velocity evolution does
conform to that of typical SNe~Ib. The depth of the 6300~\AA\
absorption of SN~1999ex was in between those of typical and
deep--H$\alpha$ SNe~Ib.  Hamuy et~al. labelled the 6300~\AA\
absorption as Si~II but this identification encounters the problem
mentioned above.  In Figure~2 a spectrum of SN~1999ex is compared with
a synthetic spectrum in which the 6300~\AA\ absorption is produced by
H$\alpha$, detached at 15,000 km~s$^{-1}$.  Undetached Ne~I
$\lambda$6402 also can fit the 6300~\AA\ absorption fairly well but
because it needs to be rather strong the other Ne~I lines cause
problems.  H$\alpha$ appears to be the most likely identification in
SN~1999ex.

\section{Helium in SNe~Ic?}

In the past, two lines of evidence have been offered for the presence
of He~I features in SNe~Ic, but both are difficult to confirm.  First,
in SN~1994I, the best observed SN~Ic (other than hypernovae),
Filippenko et~al. (1995) observed a strong absorption near 1~$\micron$
that was attributed to He~I $\lambda$10830.  However, when we (Millard
et~al. 1999) carried out a SYNOW analysis of SN~1994I, we were unable
to account for the bulk of the 1~$\micron$ absorption with
$\lambda$10830 without also producing optical He~I features that were
too strong.  We concluded that the 1~$\micron$ absorption may have
been produced at least in part by C~I $\lambda$10695 and/or blends of
Si~I lines.  Baron et~al. (1999) carried out detailed spectrum
calculations with the PHOENIX code and also had problems accounting
for the 1~$\micron$ feature in terms of $\lambda$10830 alone.

The observations of the Type~Ib SN~1999ex by Hamuy et~al. (2002)
clarify this issue.  Hamuy et~al.  obtained good infrared spectra of
SN~1999ex, and as expected from the presence of definite optical He~I
features, the infrared spectra contained P~Cygni profiles due to He~I
$\lambda$10830 and $\lambda$20581. (In LTE the Sobolev optical depth
of $\lambda$10830 exceeds that of He~I 5876 by a factor of 7 at
10,000~K and 4.5 at 15,000~K.)  However, even though SN~1999ex had
definite optical He~I lines, its $\lambda$10830 feature was weaker
than the 1~$\micron$ feature in SN~1994I, which had less distinct, if
any, optical He~I features.  It is interesting that in SN~1999ex,
unlike in SN~1994I, a SYNOW synthetic spectrum that fits the
$\lambda$10830 feature also fits the optical He~I features reasonably
well.  This supports the suspicion that the 1~$\micron$ absorption in
SN~1994I was not mainly due to He~I $\lambda$10830 --- and if it
wasn't mainly due to He~I then it doesn't provide convincing evidence
for the presence of helium at all.

The second line of evidence for He~I features in SNe~Ic has been the
identification of weak absorptions in optical spectra with
high--velocity He~I lines, detached at about 17,000 km~s$^{-1}$
(Clocchiatti et~al. 1996). However, M01 reexamined the evidence and
were unable to confirm these identifications.  One problem was that
the absorption attributed to detached He~I $\lambda$6678 usually is
substantially more blueshifted than those attributed to $\lambda$5876
and $\lambda$7065.  M01 also mentioned that the typical Type~Ib
SN~1999dn, whose deep He~I absorptions were at low velocity, contained
a weak absorption like the one that Clocchiatti et~al. attributed to
high--velocity $\lambda$5876 in SNe~Ic.  In fact, that feature often
appears in spectra of core--collapse SNe, including the Type~II
SN~1987A (Jeffery \& Branch 1990), when the temperature is low enough
for lines of singly ionized iron--peak elements to be strong.  It is
produced by a group of Sc~II lines from $\lambda$5641 to $\lambda$5684.

Being unpersuaded by the first two lines of evidence for He~I in
SNe~Ic, we consider a third.  After failing to confirm high--velocity
He~I absorptions in SNe~Ic, M01 mentioned that in the Type~Ic
SN~1990B, if He~I absorptions were present at all, they were at low
velocity.  Clocchiatti et~al. (2001) also considered the possibility
of low--velocity He~I in SN~1990B.  In Figure~3 a spectrum of SN~1990B
is compared with a synthetic spectrum that contains undetached He~I
lines.  Note that because Fe~II features are quite strong, they
account nicely for the 6300~\AA\ absorption.  He~I $\lambda$5876
accounts for an observed absorption, but of course that observed
feature could be Na~I instead.  More important is that the observed
absorption near 6800~\AA\ is accounted for by undetached He~I
$\lambda$7065, and I'm not aware of any good alternative for
undetached $\lambda$7065; neither a line of similar wavelength nor a
line of longer wavelength that could be detached and mistaken for
undetached $\lambda$7065.  An obvious problem is that there is no
counterpart in the observed spectrum for the synthetic absorption
produced by $\lambda$6678, but this is not necessarily fatal for He~I.
First, M01 showed that in SNe~Ib $\lambda$6678 gradually weakens
relative to $\lambda$5876 and $\lambda$7065; this is to be expected on
the grounds of NLTE calculations (Lucy 1991; Swartz et~al. 1993)
because $\lambda$6678 is a singlet transition while $\lambda$5876 and
$\lambda$7065 are triplets, and the singlet resonance transition to
the ground state becomes less opaque as the ejecta density decreases
through expansion.  It follows that the use of LTE excitation in SYNOW
can overestimate the strength of $\lambda$6678 relative to
$\lambda$5876 and $\lambda$7065.  Second, in SN~1990B the
$\lambda$6678 absorption would be partially filled in by the
H~II--region H$\alpha$ emission seen in the spectrum, and possibly
also by H$\alpha$ emission from SN~1990B itself (see \S6).

Another reason to take low--velocity He~I in SNe~Ic seriously is that
the early spectra of SN~1994I contained an absorption feature that we
(Millard et~al. 1999) attributed to Na~I $\lambda$5892.  But at these
early times the color temperature was high --- $\sim$17,000~K in the
spectrum obtained four days before maximum brightness.  The presence
of Na~I absorption at such a high temperature is doubtful.  In
SN~1987A, for example, Na~I began to appear only when the color
temperature fell to about 6000~K (Jeffery \& Branch 1990).  In the
early spectra of SN~1994I even the Ca~II absorptions were not yet very
strong.  The observed feature may have been produced by undetached
He~I $\lambda$5876 rather than Na~I.

During the course of an ongoing comparative study of spectra of
SNe~Ic, similar to that of B02 for SNe~Ib, it has become clear that
absorptions that can be attributed to undetached He~I $\lambda$5876
and $\lambda$7065 are common in the early spectra of SNe~Ic, and
sometimes a weak absorption consistent with $\lambda$6678 also is
present.  


\section{Hydrogen in SNe~Ic?}

In the past, two lines of evidence for the presence of H$\alpha$ in
SNe~Ic also have been presented, but again confirmation is difficult.
Filippenko, Porter, \& Sargent (1990) and Jeffery et~al. (1991)
tentatively attributed a weak absorption near 6360~\AA\ in SN~1987M to
H$\alpha$, but as emphasized by Wheeler et~al. (1994) the required
blueshift of H$\alpha$ was suspiciously low compared to those of other
lines in the same spectrum. In any case, the absorption in question
is unusual, perhaps even unique so far; it is not the same as the
broader feature that is commonly referred to as the 6300~\AA\
absorption, which ordinarily appears at a wavelength roughly
consistent with undetached Ne~I $\lambda$6402 (whether Ne~I is the
right identification or not).  The 6360~\AA\ absorption in SN~1987M is
farther to the red.

More intriguing is that Filippenko (1988, 1992) pointed out that
SNe~Ic have a broad emission peak near the wavelength of H$\alpha$.
We must be careful about attributing this feature to H$\alpha$
emission when the Fe~II features are strong --- but it appeared even
in the pre--maximum spectra of SN~1994I, when the Fe~II features were
not yet strong. It is indeed tempting to attribute this feature to
H$\alpha$.  The H$\alpha$ would need to be in net emission (as it
often is in SNe~II), and to produce the rounded emission peak the
H$\alpha$ would need to be undetached.

\section{Discussion}

So --- {\sl do SNe~Ib eject hydrogen?}  Yes, the deep--H$\alpha$
events certainly do; they appear to be closely related to SNe~IIb.  In
fact, the differences between SNe~IIb and the deep--H$\alpha$ SNe~Ib
can be so small that the classification of some events probably
depends on how early the first spectrum is obtained.  [In the
nomenclature of Clocchiatti \& Wheeler (1997) the deep--H$\alpha$
events could be included among the ``transition supernovae'',
SNe~IIt.]

Whether typical SNe~Ib eject hydrogen is much more difficult to
decide.  Given the strong similarities between the deep--H$\alpha$ and
typical SNe~Ib (B02), the most natural assumption would seem to be
that the 6300~\AA\ absorption in typical SNe~Ib also is produced by
H$\alpha$.  This is supported by the probable presence of
intermediate--strength 6300~\AA\ absorption in the shallow--helium
SN~1999ex.  On the other hand, nonthermally excited Ne~I lines are not
implausible, and because Ne~I $\lambda$6402 would not need to be
detached it would require one fewer free parameter (the detachment
velocity of H$\alpha$).  The 6300~\AA\ absorption tends to drift
redward with time like the other features; this may seem to favor
undetached Ne~I, but it is not decisive since the detached He~I lines
in SNe~Ib also drift to the red.  (B02 offer an explanation.)  How are
we to decide about the H$\alpha$ identification?  The most direct
confirmation would be to detect Paschen--alpha
(P$\alpha~\lambda$18751), but in LTE its Sobolev optical depth is
lower than that of H$\alpha$ by a factor of 34 at 5000~K and 3.4 at
10,000~K, so this may not work.  A larger sample of high quality
optical spectra of typical SNe~Ib will allow the following test to be
applied: if the 6300~\AA\ feature {\sl always} is consistent with
undetached Ne~I, then it probably is Ne~I, because it would be hard to
believe that hydrogen always is detached by the same amount with
respect to the velocity at the photosphere.  The current data are
inconclusive on this point.

{\sl Do SNe~Ic eject helium?}  As we have seen, the 1~$\micron$
feature in SN~1994I does not provide convincing evidence for He~I
$\lambda$10830, and the presence of weak optical absorptions with
highly detached He~I lines is unlikely.  However, it may be that the
optical spectra contain undetached He~I lines.  How are we to decide?
Since He~I $\lambda$10830 can be confused with lines of other ions,
the most direct test might be to look for $\lambda$20581, which in LTE
has about the same Sobolev optical depth as $\lambda$10830 above
10,000~K.  (But be careful --- $\lambda$20581 is a singlet
transition.)  Infrared spectra also could be used to sort out the
contributions of C~I and Si~I to the 1~$\micron$ feature because both
ions have other strong lines that should be detectable (Figure~12 of
Millard et~al. 1999).

{\sl Do SNe~Ic eject hydrogen?}  Identifying the weak 6360~\AA\
absorption in SN~1987M with H$\alpha$ may be questionable, but the
possibility of emission near the rest wavelength of H$\alpha$ looks
interesting. The hydrogen would need to be undetached.  This, together
with the possibility that SNe~Ic contain undetached He~I lines, means
that if SNe~Ic do eject helium and hydrogen then these elements extend
to {\sl lower} ejection velocities than in SNe~Ib. The appearance of
the hydrogen and He~I features in SNe~Ib and SNe~Ic may be affected by
the radial distributions of the helium and hydrogen in addition to the
total ejected masses of helium and hydrogen.  We should consider the
possibility that the composition structures of SNe~Ic are more mixed
up than those of SNe~Ib.

This discussion of line identifications has been based on calculations
carried out with the simple SYNOW code.  Detailed NLTE calculations
for a grid of explosion models for SNe~Ib and SNe~Ic are needed,
especially to test the possibilities of nonthermally excited Ne~I
lines in SNe~Ib (and SNe~Ic), and of rounded H$\alpha$ emission in
SNe~Ic.  It also is needed in order to estimate, or place upper limits
on, the masses of helium and hydrogen in SNe~Ib and SNe~Ic.

Clocchiatti et~al. (1997) have emphasized that light--curve shapes
imply significant differences in the ejected masses of supernovae of
the same spectral type.  This, together with our uncertainty about the
composition structures of the ejected matter, and our still scanty
knowledge of the relative occurrence frequencies of the various kinds
of SNe~Ib and SNe~Ic, means that there is much work to be done before
we will be able to match the various kinds of SNe~Ib and SNe~Ic with
their stellars progenitors in a detailed way.

\acknowledgements

I am grateful to Tom Matheson and Mario Hamuy for providing spectra;
to them, Eddie Baron, Jinsong Deng, Chris Gerardy, and David Jeffery
for comments on the manuscript; and to members of the University of
Oklahoma supernova group for many discussions.  This work has been
supported by NSF grant AST--9986965.

\begin {references}

Baron, E., Branch, D., Hauschildt, P. H., Filippenko,~A.~V.,
Kirshner,~R.~P. 1999, ApJ 527, 739

Benetti, S. et al. 2002, MNRAS in press; astro-ph/0205453

Branch, D. 1972, A\&A 16, 247

Branch, D. et al. 2002, ApJ 566, 1005 (B02)

Branch, D., Baron, E., Jeffery, D. J. 2002, in: K.~W.~Weiler (ed.),
Supernovae and Gamma--Ray Bursts (New York: Springer), in press;
astro-ph/0111573

Clocchiatti, A. et~al. 1997, ApJ 483, 675

Clocchiatti, A. et al. 2001, ApJ 553, 886  

Clocchiatti, A., Wheeler, J. C. 1997, ApJ 491, 375

Clocchiatti, A., Wheeler, J. C., Brotherton, M. S., Cochran,~A.~L.,
Wills,~D., Barker,~E.~S., Turatto,~M. 1996, ApJ 462, 462

Deng, J., Qiu, Y. L., Hu, J. 2002, ApJ submitted; astro-ph/0106404

Deng, J., Qiu, Y. L., Hu, J., Hatano,~K., Branch,~D.  2000, ApJ
540, 452

Filippenko, A. V. 1988, AJ 96, 1941

Filippenko, A. V. 1992, ApJ 384, L37

Filippenko, A. V. 1997, ARA\&A 35 309

Filippenko, A. V. et al. 1995, ApJ 450, L11

Filippenko, A. V., Porter, A., Sargent, W. L. W. 1990, AJ 100, 1575

Hamuy, M. et al. 2002, AJ, 124, 417

Harkness, R. P. et al. 1987, ApJ 317, 355

Jeffery, D. J., Branch, D., 1990, in: J.C.~Wheeler, T.~Piran, \&
S.~Weinberg (eds.), Supernovae, Sixth Jerusalem Winter School for
Theoretical Physics (Singapore: World Scientific), 149

Jeffery, D. J., Branch, D., Filippenko, A. V., Nomoto,~K. 1991,
ApJ 317, 717

Kurucz, R. L. 1993, CD-ROM 1, Atomic Data for Opacity Calculations
(Cambridge: Smithsonian Astrophysical Observatory)

Lucy, L. B. 1991, ApJ 383, 308

Matheson, T. et~al. 2000, AJ 120, 1487

Matheson, T., Filippenko, A. V., Li, W., Leonard,~D.~C.,
Shields,~J.~C. 2001, AJ 121, 1648 (M01)

Millard, J. et al. 1999, ApJ 527, 746

Qiu, Y., Li, W., Qiao, Q, Hu, J. 1999, AJ 117, 736

Swartz, D. A., Filippenko, A. V., Nomoto, K., Wheeler,~J.~C. 1993,
ApJ 411, 313

Turatto, M. 2002, in: K.~W.~Weiler (ed.),  Supernovae and Gamma--Ray
Bursts (New York: Springer), in press

Wheeler, J. C., Harkness, R. P., Clocchiati, A., Benetti,~S.,
Brotherton,~M.~S., DePoy,~D.~L., Elias,~J. 1994, ApJ 436, L135

\end{references}  

\clearpage

\begin{figure}
\plotfiddle{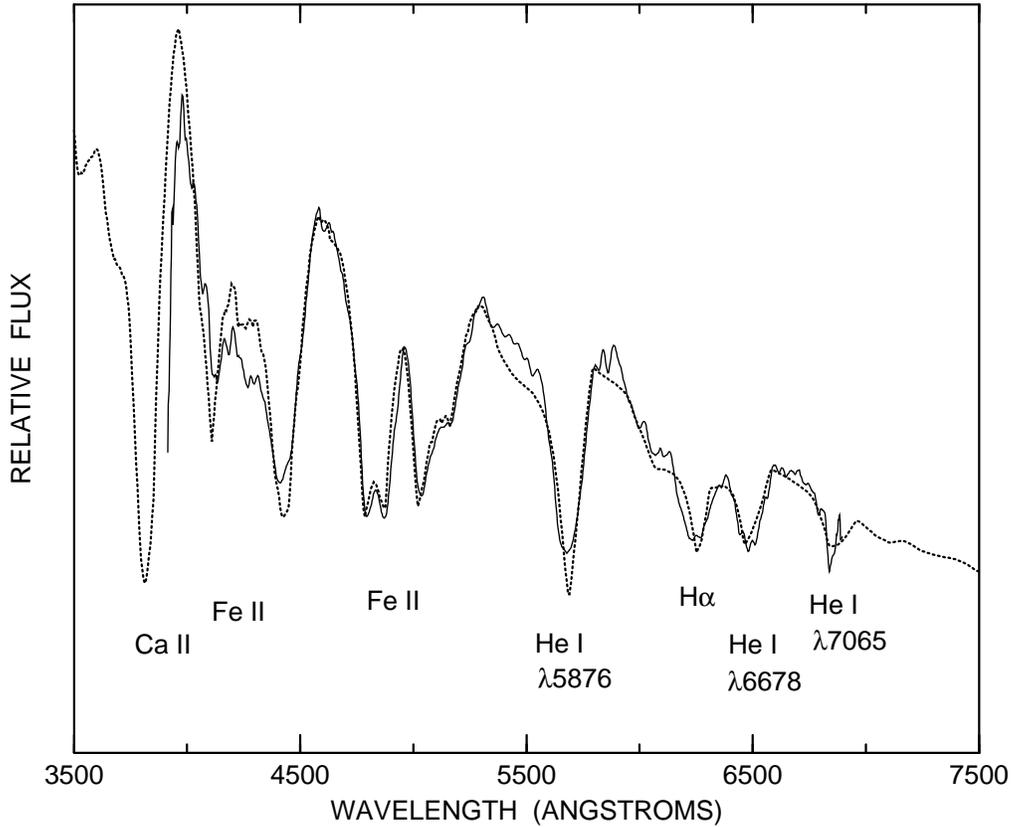}{3cm}{-90}{70}{70}{-250}{375}
\caption{A spectrum of the Type~Ib SN~1984L obtained by Harkness
et~al. (1987) 9 days after maximum brightness (solid line) is compared
with a synthetic spectrum (dotted line) that has $v_{phot}= 9000$
km~s$^{-1}$ and contains lines of H, He~I, Ca~II, Sc~II, and Fe~II.
The hydrogen lines are detached at 15,000 km~s$^{-1}$.  In this and
subsequent figures the flux is per unit wavelength interval and its
scale is arbitrary. Figure adapted from Branch et~al. (2002)}
\end{figure}

\clearpage

\begin{figure}
\plotfiddle{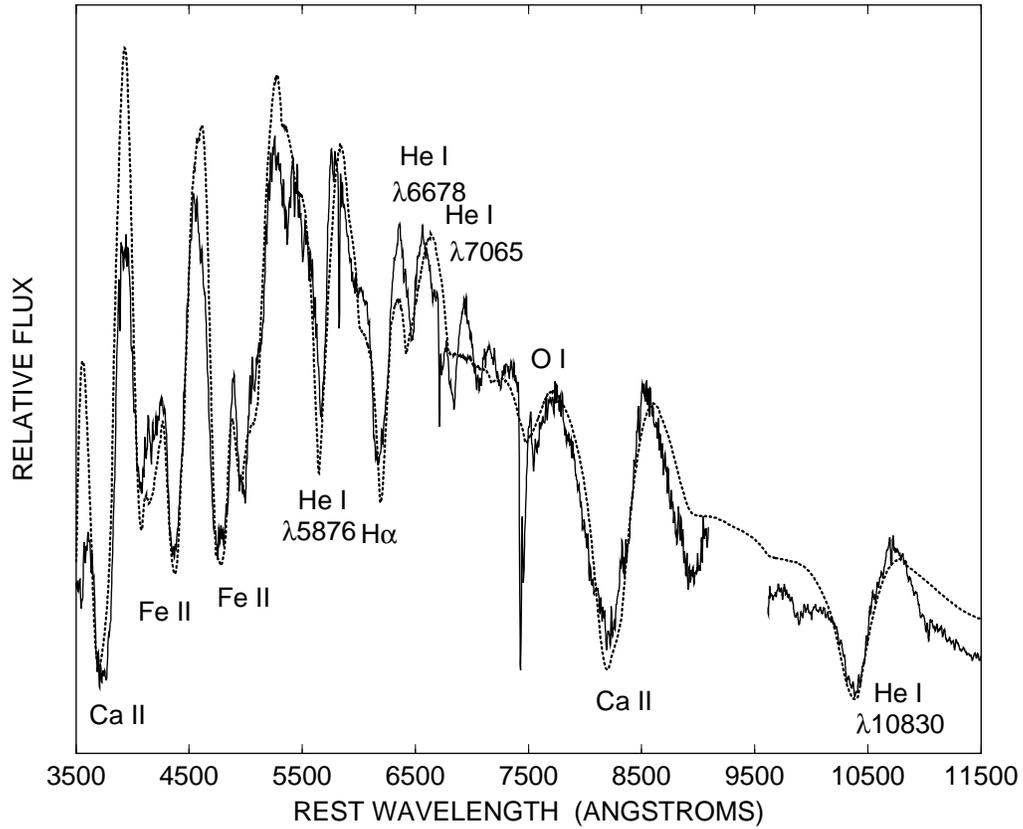}{3cm}{-90}{70}{70}{-250}{375} 
\caption{A spectrum of the Type~Ib SN~1999ex obtained by Hamuy
et~al. (2002) 4 days after maximum brightness (solid line) is compared
with a synthetic spectrum (dotted line) that has $v_{phot}= 9000$
km~s$^{-1}$ and contains lines of H, He~I, O~I, Ca~II, and Fe~II.  The
hydrogen lines are detached at 15,000 km~s$^{-1}$.}
\end{figure}

\clearpage

\begin{figure}
\plotfiddle{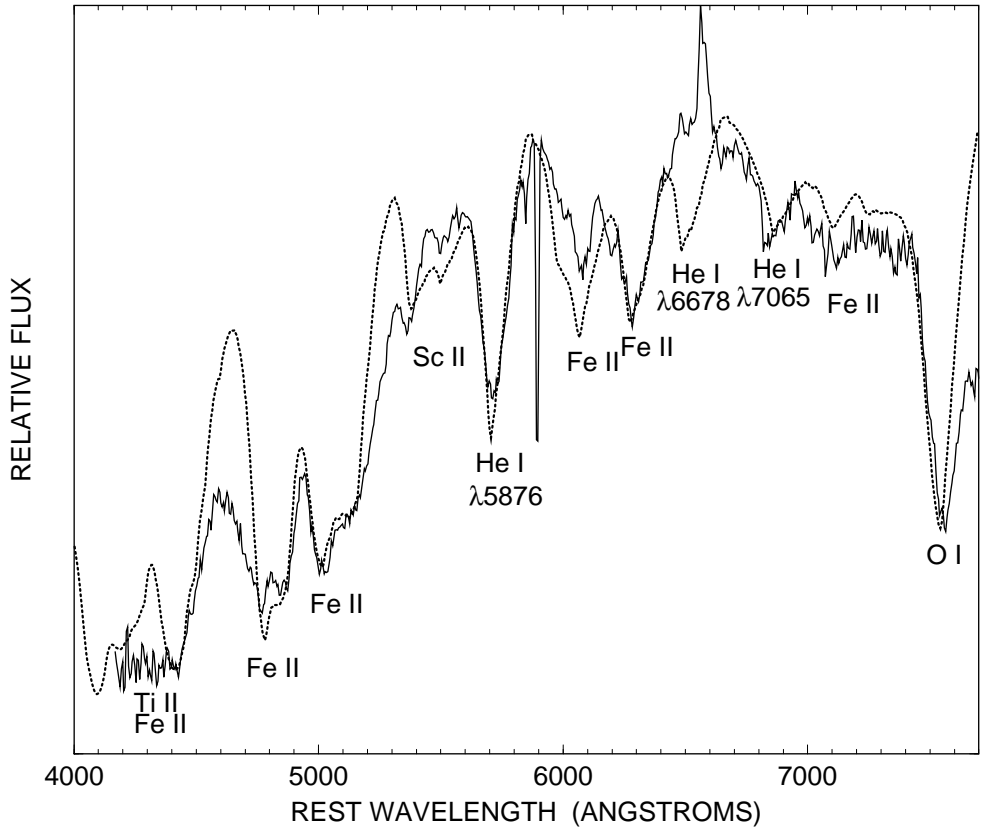}{3cm}{-90}{70}{70}{-250}{375}
\caption{A spectrum of the Type~Ic SN~1990B obtained by Matheson
et~al. (2001) 5 days after maximum brightness (solid line) is compared
with a synthetic spectrum (dotted line) that has $v_{phot}= 9000$
km~s$^{-1}$ and contains lines of He~I, O~I, Ca~II, Sc~II, Ti~II, and
Fe~II.  The narrow peak at 6563~\AA\ is H$\alpha$ emission from an
H~II region and the narrow absorption at 5892~\AA\ is due to
interstellar Na~I.}
\end{figure}

\end{document}